# Real-Time Image Processing Algorithms for Embedded Systems

SOUNDES OUMAIMA BOUFAIDA [1], ABDEMADJID BENMACHICHE [1] and MAJDA MAATALLAH [1]

[1] Department of Computer Science, LIMA Laboratory, Chadli Bendjedid University, El-Tarf, PB 73, 36000, Algeria

*E-mails*

*s.boufaida@univ-eltarf.dz*; *benmachiche-abdelmadjid@univ-eltarf.dz*;

*maatallah-majda@univ-eltarf.dz.*

**Abstract**

Embedded vision systems need efficient and robust image processing algorithms to perform real-time, with resource-constrained hardware. This research investigates image processing algorithms, specifically edge detection, corner detection, and blob detection, that are implemented on embedded processors, including DSPs and FPGAs. To address latency, accuracy and power consumption noted in the image processing literature, optimized algorithm architectures and quantization techniques are employed. In addition, optimal techniques for inter-frame redundancy removal and adaptive frame averaging are used to improve throughput with reasonable image quality. Simulations and hardware trials of the proposed approaches show marked improvements in the speed and energy efficiency of processing as compared to conventional implementations. The advances of this research facilitate a path for scalable and inexpensive embedded imaging systems for the automotive, surveillance, and robotics sectors, and underscore the benefit of co-designing algorithms and hardware architectures for practical real-time embedded vision applications.

## 1. Introduction

This essay presents a comparative study of some image processing algorithms in embedded vision systems which have been benchmarked using real-time recording and artificial image datasets. Low-density parity check and motion vector algorithms detect inter-frame redundancy in every processed field to achieve high frame-rate coded output. Intra-frame prediction filters block-derivable DC coefficients with a residual correction of 2D frequency components to reduce Coded Video Discussion data without compromising picture quality. Intra-field coding, frame-averaging filtering combined with an inter-field alternate macro-block transfer code improves transmission frame rate while reducing coding latency.

Design guidelines are derived to deploy the image processing tasks on a multi-field basis while maintaining data and picture quality robustness. Timing budget gives a better estimation of the maximum number of admissible fields which can be processed in PLDC algorithm, making it possible to dimension the PLDC devise accordingly. Picture analysis gives an optimal combination of the other algorithms, achieving ratio results for transmitted frames in cooperation. It is shown that those algorithms can be used independently without compromising their function. Subsequent coded video presentation gives a comprehensive and illustrative view of the video sequences before and after being processed by these algorithms. Fixed point simulation analysis of the deployed algorithm in the target architecture shows that picture quality robustness is preserved.

Embedded Vision Systems capturing, treating and transmitting images is a critical function in the automotive domain. Pioneer work pioneered it by formulating image compression, transmission and graphical user interface (GUI) rendering in the standard C programming language. Previous works developed versatile image processing algorithms predicting inter-frame redundancy, basing the treatment on integral z-transform. Some algorithms were adapted to hardware implementation on a Xilinx Virtex-II field programmable gate array (FPGA) with embedded microprocessor. This ensured in real time processing compliance with the ISO 26262 hardware safety integrity level (ASIL) requirements. Most of these algorithms were developed considering a single frame basis and/or could not run complementary to each other. Recent studies point to the necessity of creating adaptive scheduling schemes as well as learning-based perception models in embedded vision systems for controlling time-sensitive execution constraints and improving system robustness in complex application situations [1], [2], [3].

## 2. Background of and Motivation

Real time image processing systems are the systems which check for conditions at regular time intervals. If there is any deviation with respect to the condition, then appropriate action will be taken. Real time image processing systems can be used for detecting vices or monitoring places like banks, ATM centers, Entry and exit of sensitive places like laboratories, etc[4]. There are many embedded systems which process the image in digital or analog form but cannot process in real time. There is a need for embedded systems with image processing capable of processing the images in real time.

## 3. Scope and Objectives

The Scope and Objectives of the essay are outlined in this section. It provides clarity on the boundaries within which the topic is explored and what specific goals are aimed at being achieved.

### 3.1. Scope

Monitoring of surveillance places has become a need recently, and this had to be fulfilled with other embedded image processing systems. Also, the images captured in these systems should be processed in real-time for efficient performance. There are various algorithms that are designed and standardized for image processing. The various image processing algorithms such as edge detection, object recognition, etc., are implemented here using the C81xx processor (525 MHz) from Texas Instruments [5].

This processor has many features, which provide dedicated DSP features as well as video enhancement features suitable for image processing. A visual board is designed around this processor with all required peripherals such as display and camera to achieve the project objectives. The performance of the different algorithms is compared with the fixed point and floating-point processing.

### 3.2. Objectives

This section will serve as an introduction to the product. The intention is to stimulate interest in it and give an overview of its most important features [1]. More detailed information will be supplied in the body of the report, including application contexts and background theories. Image processing is one of the efficient tools for enhancing the captured image. There are many image enhancement techniques available for enhancing and processing the images. The latest DSP processors are being used to overcome the difficulty of processing time and arrangement of the hardware. Addition and difference of the images, various filters, morphological operations, data compression schemes, etc. are most widely used techniques in the existing image processing systems.

## 4. Fundamentals of Real-Time Image Processing

This section serves as a foundation for understanding real-time image processing. Basic concepts and principles essential for comprehending subsequently advanced topics of this research area are covered. The first half of the section provides an overview of the basic concepts of images, digital images and images in their mathematical representation. The second half deals with basic image processing concepts like filtering, enhancement and reconstruction. The real-time image processing basics presented in this section provide a good foundation to understand the complex topics of vision algorithms like stereo vision, optical flow, motion tracking, object recognition, pattern matching, active vision etc [6].

### 4.1. Image Processing Basics

An image is a two-dimensional signal (or three-dimensional if colors are included) received from some physical scene[7]. Spatial imagery formed through the projection of scenes is studied for improving and extracting the data contained in the images. Typical applications exist in areas such as video surveillance, remote sensing, medical imaging, and non-destructive testing. Image processing is generally grouped into two broad categories: conventional techniques and real-time image processing techniques [8]. Such applications include the restoration of medical samples before diagnosis, improving image quality in remote sensing scenarios, and decoding and displaying images received from low-resolution cameras mounted on satellites or Simple Observation Platforms around the earth. Many pixels need to be processed in a short time to make the above task viable. Real-time implementation of image processing algorithms is extremely challenging since the hardware resources of typical image processors are very limited.

### 4.2. Real-Time Constraints

Real-time processing is the ability to keep up with the incoming data stream. When discussing real-time, there is often confusion about the different flavors of real-time. There are hard real-time systems that cannot tolerate delays and will cause catastrophic failures, such as flight control systems for aircraft. There are soft real-time systems that may fall behind, but can tolerate some missed frames, such as multimedia streaming. In fact, the terms hards and soft real-time have more to do with business decisions made than with the data stream itself. When discussing real-time processing, it is useful to provide an example of data that may be processed.

The previous sections focused mostly on image processing. It is common, especially in the scientific community, to be beset by the image. However, images are just one data stream and there are several others that can be used and been used effectively.

Both data streams have hard real-time constraints on them. This can be best categorized by describing what happens if the processing does not keep up. For images, they build up in a buffer waiting to be processed, adding lag to the system. For audio, if the processing cannot keep up the data is simply discarded. In the worst case, there will be terrible noise that is unlistenable. Understanding the response time requirements for a data stream is imperative to designing a real-time system capable of processing that data stream [9]. The underlying hardware for anything more complicated than trivial processing will be vastly different depending on the data stream.

## 5. Fast Fourier Transform in Real-Time Image Processing

The DFT relates a sampled function in the time domain to its sampled function in the frequency domain. It is a complex-valued summation requiring $O(n^2)$ multiplications and additions, with n the number of samples. It is linear, periodic, and can be viewed in the frequency domain as the transfer function of a filter bank. The fast Fourier transform (FFT) computes the DFT in $O(n \log n)$ time using the divide-and-conquer paradigm.

### 5.1. Overview of FFT

For many applications in communication systems, medical imaging, navigation, process control, remote sensing and in many others, real-time image processing is of great interest. In such applications, filtering algorithms such as convolution filter or correlation filter are computationally expensive which involves large number of multiplications and additions between the pixels of the image and filter coefficients. Filtering operations involving large number of pixels in the image and large filter coefficients are implemented efficiently in the frequency domain, where filtering operations require fewer number of computations. To convert image from spatial domain to frequency domain and vice versa, Fourier Transform is used [10]. Basic Fourier Transform is highly complex algorithm and has large time complexity which renders it unsuitable for real-time applications and so a FFT algorithm is required. Basically, there exists a trade-off between speed and hardware requirement of FFT algorithms. In general, FFT processors fall under two designs i.e. high-speed design architectures where throughput is increased at the cost of increased hardware resources and scalable architectures where throughput is increased at the cost of reduced hardware resources [11]. Thus, single-path delay feedback (SDF) multi rate FFT processors can be used to address the applications which require high speed with minimal hardware resources.

### 5.2. Applications in Real-Time Image Processing

The two-dimensional discrete Fourier transform (2D DFT) is important for many applications including image processing, pattern recognition and machine vision. Fourier image analysis has been used to reduce computational complexity of many complex convolution operations involving derivative operators in the spatial domain by converting them into simple multiplications in the frequency domain [12]. Nevertheless, as a direct consequence of using DFTs, large amounts of images that are required for processing can lead to bottlenecks in machine vision applications necessitating high-speed parallel architectures. In addition to being memory constrained, DFTs in many systems form bottlenecks in the control loop of the machine vision tasks, making it essential to accelerate their computations. This is particularly the case for standard image sizes such as those from the common CCIR or PAL cameras with a resolution of 512×512. On such images, the computational requirements of 2D DFT are inherently high. It is well known that the Cooley-Tukey FFT algorithm reduces the computational complexity of N-point DFTs from $O(N^2)$ to $O(N \log_N)$. However, in the case of 2D DFTs, it has been shown that the implementation of 1D FFTs must be computed in two dimensions, i.e., 1D FFT must be implemented before and after the row-to-column data transposition. This increases the computational complexity of the design, in addition to the control and management circuitry, making 2D DFTs a significant computational bottleneck for many real-time machine vision applications. There exist several resource-efficient, high-throughput implementations of 2D DFTs. However, the majority of the currently available on chip FPGA based 2D FFT FPGA implementations rely on a repeated invocation of 1D FFTs by the row and column decomposition (RCD) [13]. RCD schemes have been proposed and hardware architectures described to achieve for the most used PPUs, fully pipelined with parallel processing and several very promising systems achieving real-time or near real-time performance upon the standard image such as the previous examples. One significant complexity issue/challenge with RCD schemes is the transposition switch which is often very difficult to design efficiently due to the several complex back-to-back multiplexers. In practice, all RCD-based designs either partially or entirely avoid all-to-all data transposition.

## 6. CNNs in Real-Time Image Processing

CNNs are composed of layers, where each layer consists of a few filters (also referred to as convolutional kernels). These filters are trained during the training phase and are employed by the network to perform image processing operations like edge detection, texture filtering and so on [14]. CNNs can be employed to classify an image at a broader level (semantic labels of the image), while models known as Fully Convolutional Networks (FCNs) carry out pixel classification tasks (instance level labeling). CNNs and FCNs have been shown to yield state-of-the-art performance in various high-level vision applications such as object detection and recognition in images and videos, semantic segmentation, and artistic style transfer [15], [16].

However, implementing computer vision algorithms based on CNNs and/or FCNs poses critical challenges for Embedded Systems, which need to simultaneously ensure real-time processing and power consumption at a given threshold [17], [18]. Two broadly defined challenges can be identified here: 1) Network complexity and resource consumption, and 2) Input image complexity. In benchmark datasets like ImageNet, videos to be processed have a resolution of 227 × 227 pixels and only three-color channels. Since cameras in Embedded

Systems usually output high-resolution color images (e.g., > 1 megapixel), there is thus a compelling need for optimizing neural networks tackling image processing tasks. Therefore, real-time video/image segmentation and processing with high-resolution images on embedded platforms remain an open challenge [19].

### 6.1. Introduction to CNNs

CNNs are an extensively employed deep learning architecture, which allows the automatic extraction of multi-scale image features without requiring prior knowledge to be integrated by users [20],[21]. A CNN has four main building blocks, including a convolution layer, a pooling layer, a fully connected layer, and an output classifier [22]. A convolution layer has M filters, which are trained offline and passed to the architecture. Each filter is a 3D volume with ($3×3×K$) weights (three for each RGB channel) and one bias. The result of a convolution is a joint convolved image, which indicates how strongly each filter at a certain position responds to the input image. Pooling layers are aimed at reducing the dimensionality and robustness to allowable image deformation. Therefore, the scale-invariance property of pooling layers reduces the representation size while maintaining the maximum important information. The pooling operation is performed after the convolution between the input image and filter responses, where the pooling size determines how much of shrinking size the operation would have. Each pooling layer can be followed by several convolution layers to decrease the picture size gradually. A fully connected layer performs a matrix multiplication between the vectorized response of the last layer and the trained $M×N$ weights, where N indicates the number of units in the fully connected layer. Each fully connected layer is often swanned by a rectified linear unit (ReLU) activation function to allow non-linearity.

CNN features are a function of the learned layer and are expected to be selective for certain local aspects of the input data. The deeper the layer is, the higher-level feature is, and the more complex the learned patterns are. The estimation of features based on CNNs has been widely adopted in earlier works [23].

### 6.2. Optimization Techniques for Embedded Systems

Various optimization techniques and strategies, tailored for the deployment of CNNs to embedded platforms are presented in Holistic Optimization of Embedded Computer Vision Systems [24]. This covers solutions specifically addressing challenges of real-time image processing, a central capability for autonomous systems, across vision sensors from programmable and heterogeneous SoCs to low-power FPGAs and imager with processor co-designing. Prototyping, benchmarking and model deployment workflows across neuron abstraction-levels from low-precision fixed-point weights to tiny-LAT-based models and hardware frameworks are also described. These techniques can significantly reduce a CNN's computational complexity and memory requirements. Methods for quantizing a model's weights and activations to 8 bits or lower fixed-point with proven robust performance, forming fast and area-efficient designs for programmable platforms are demonstrated. Memory access requirements can be reduced by converting CNNs into two-stream architectures where activations at computation-levels are reused on-chip and redundant computations are eliminated[25]. Further reductions in area and latency are achieved through the pruning of weights and neurons of the models with respect to pre-trained performance, resulting in designs comparable within reach of hardware framework's RAM budget. To improve model mapping on programmable SoCs and heterogeneous systems with accessible memory hierarchies and IP cores, multi-dimensional scheduling methods are also developed. These algorithms aim to maximize performance on embedded hardware with fixed performance metrics through the orchestration of data transfer, access pattern and computation.

## 7. Parallel Computing Techniques in Real-Time Image Processing

The ability to parallelly process data has gained importance in modern image processing applications. Multiple processors can be used to manipulate large images, which would otherwise be cumbersome when processed by a single processor. Heavy processing loads are increasingly imposed on different embedded systems as imaging devices with larger resolutions become available. Image acquisition rates in today's cameras are also on the rise. Processing images in real time is key to many applications. This imposes stringent requirements on speed to processors which may deal with large image databases. It has also become necessary to make various algorithms computationally simpler. Parallel processing is a possible solution [26],[27], which achieves more speed by conducting some operations simultaneously. Nevertheless, the extent to which this would be beneficial for real time processing based on specific applications needs careful analysis.

To study the possible gains available from parallel processing in image processing applications, typical algorithms for image filtering and transformation are examined. The degrees of parallelism achievable with these algorithms on ICO-V and other parallel computers are examined. Some strategies to be followed for processing images in parallel are also pointed out. The parallel image processing algorithm is constructed after considering the strategies and the architecture of an ICO-V parallel computer. Experimentally attainable speedups for this case differ very much from theoretical speedups. Some performance results are presented. Close to linear speedup can be expected with increasing number of processing elements for image frame filtering when square filter masks are used. However, the degree of parallelism diminishes with increasing filter mask size. The presence of large numbers of pixels with a given intensity may cause the speedup for binary image processing based on decision trees to approach a constant value even in a very highly parallel architecture [9].

## 7.1. GPU Acceleration

As the number of pixels and cameras in an image processing event gets higher, the parallel processing ability of a system becomes vital for ad-hoc real-time performance. In recent years, the use of Graphics Processing Units (GPUs) provides an ability of high computational throughput, capable of processing 1000s to millions of paralleled tasks at once. The massive parallel processing architecture of GPU brings significant advantages for Real-Time Image Processing Applications when there are lots of processing units with simple math operations in image data. There are several architectures available for embedded systems equipped with GPUs. In this paper, an experimental study is accomplished to compare the performance of OpenCV built-in CPU and GPU functions on a Cortex A8 Nvidia Tegra 2 based embedded platform. In addition, the performance on a quad-core CPU platform is investigated for better understanding the behavior of processing algorithms on parallel architectures [9].

Managing a vast processing capability is not only dependent on the architecture of the supporting system but also the algorithmic performance on the system is quite essential. In a limited time, frame of an image sensor event, an out-sourced algorithm must be thought fast enough to handle the real-time requirement. The ratio of the time to run a particular process with a particular algorithm, versus the ratio of the time is consumed to run the same process with another algorithm is called algorithmic performance. It is expected to build self-compliant embedded versions of the computationally intensive Image Processing Applications designated to AmpliSens Processor, NVIDIA's low power Quad-core ARM Cortex A57 generation architecture [28].

## 7.2. Parallel Processing on FPGAs

There is a focus on parallel processing for real-time image processing using Field-Programmable Gate Arrays (FPGAs). Increasing demand for high-speed processing results in a need for parallel processing. In the case of image processing applications, it is usually very complex and requires many operations. Hence, a large amount of data is transferred which requires high-speed data transfer. Therefore, multiprocessing or parallel architecture is needed to cope with this situation. Multi-tasking operating systems and bus architectures can provide parallel processing but not beyond a limit because of the serial nature of data transfer. Therefore, some hardware in which multiple processing cores can be used simultaneously is preferred. In this regard, FPGA helps with parallel processing, and hence a better alternative for real-time image processing. It consists of logic cells, I/O pins, interconnects, and programmable interconnects. The entire FPGA architecture can be implemented from HDL (Hardware Description Language) code and is very flexible and supports programmability. Since FPGAs have a dedicated path for data transfer, they also provide a rapid data rate [6]. Interconnects between IOs and logic cells are also programmable. Each IO can be configured with phase-lock loops and series resistors, making it flexible to work with high/low-speed interfaces. So, implementation of FPGA for image processing is preferred over other platforms. There are many successful image processing applications like it's hard to imagine an application that needs high-speed processing and can't be applied to DSPs. Prerecorded/streaming video can be easily interfaced with low-cost chipsets [17].

## 8. Performance Evaluation and Comparison

Performance evaluation is an important consideration when developing real-time image processing algorithms. This consideration influences the design, choice and employed methodologies to achieve the desired performance criterion. On Embedded Systems (ES) design, performance evaluation can be a critical aspect of the overall design as it impacts the architecture of the chosen hardware and/or algorithm implementation. If real-time behaviour is desired, the performance evaluation of algorithms is typically necessary to understand performance limits of a system [9]. The metrics applied to determine performance criteria can take-on several names, such as Performance Parameters, Performance Evaluation Metrics or Performance Measures, which characterise criteria used to determine performance, which may influence future design decisions.

The envisioned real-time image processing algorithms are usually compared to traditional image processing algorithms [17]. Comparisons between real-time Dehazing, Denoising, Inpainting, Math Morphology and Resizing approaches against traditional image processing approaches which have close application to the within-outside pixel whether the value of a pixel is calculated and processed if it is different from that of its neighbors. A set of metrics to measure image digitised distortion may be used (e.g., PSNR, MSE, Maximum Absolute Difference, number of pixels out of tolerance).

### 8.1. Metrics for Evaluation

The performance of each defined algorithm is evaluated according to the following metrics:

- Latency and throughput are measured directly from the embedded platforms. Frames per second (FPS) is calculated from the observed processing latency, and throughput is expressed as pixel operations per second (POP/s) [9].
- Accuracy is evaluated according to the error rate for the associated image processing application, namely for corner detection, template matching, and optical flow. The detection accuracy (detection rate) is derived from the number of correctly detected features with respect to the number of features in the reference images, and the matching accuracy is derived from the matching error, which is computed from the squared Euclidean distance between matched feature descriptors [5]. Performance characteristics are analyzed using speedup and efficiency. Speedup is defined as the ratio of the performance of the faster processor or configuration to that of the slower one (using FPS). Efficiency is defined as the speedup divided by the number of processors or cores used in parallel processing.

## 8.2. Benchmarking Studies

To understand the performance of the systems developed experimentally, several benchmarks Realtime image processing algorithm implementations were carried out on the systems. The image processing algorithms chosen were widely used in various computer vision applications and widely researched in the image processing literature. Some of these algorithms have relatively simple data flow such as thinning, edge detection and histogram operations [29],[30]. Other algorithms have a more complicated data flow such as mesh grid generation and watershed segmentation which are also more computationally intensive. In addition to the image processing algorithms, the image filtering operation used to generate the input image data sets was also chosen for benchmarking because it involves convolution which has a complicated data dependency and is commonly used to preprocess images for edge detection [6]. As a preliminary analysis the algorithms and image filter kernel files used for the analysis were applied to a few different image sizes, 8-bit grey level images only. The image processing operations chosen were formed together to form a common data flow of operations that were simulated progressively one after the other [31]. Timing results were generated for each image processing operation individually and collectively to understand how each operated and additionally how each affected the other. These timings give an indication of the average true processing time required for each image processing operation on the different image sizes. This information can then be further developed in the safety of a simulation environment to understand the capabilities of the processors used in potential application systems.

# 9. Challenges and Future Directions

The field of real-time image processing has witnessed significant developments in recent years; however, it still faces considerable challenges in the design of algorithms and the hardware implementation of solutions and systems. Many attempts are being made to apply image processing and computer vision concepts for implementation on embedded platforms. Today, advents in custom hardware, reconfigurable hardware, high-speed chip designs, low-cost components, power-full microcontrollers and DSP architectures, image data compression solutions, and high-speed interfacing protocols provide every opportunity to design systems and solutions that can work standalone, require low bandwidths, and consume low power. Applications that require high reliability under unforeseen problems, such as with respect to illumination changes and variations in the scene, deem smart processing on-chip, in a vision system, as opposed to computationally intensive tasks executed by off-chip general-purpose hardware.

With the increasing time and effort required for systems integration, system-on-chips, and hardware-in-the-loop simulations are becoming more popular. The road towards product realization typically starts with a reference implementation on several readily available platforms, such as EDA software for the high-level modeling of systems, FPGAs for prototyping, and higher-level development frameworks on DSP and application-specific processors. Moreover, rapid bandwidth growth develops new possibilities for connection architectures, allowing us to make quick hardware enhancements, e.g., by integrating data buses dedicated to a newly implemented vision module [3]. The main driver in this direction is the fast-growing sector of mobile vision applications that imposes low manufacturing and processing costs in combination with high-performance and real-time-processing demands. Therefore, new and smart solutions that allow not to compromise performance on existing high-end platforms need to be investigated. Also, in most recent research, it was shown that developing an integrated approach of smart security and privacy-preserving solutions along with resilient systems architecture will become a major challenge for future real-time embedded vision systems operating in connected and sensitive data environments [32], [33] [29].

## 9.1. Current Challenges

Real-time image processing applications, such as surveillance, robotics, tracking, and gesture recognition, are gaining popularity due to the proliferation of camera sensors. Embedded systems, which are cost-effective alternatives to standard computers, using a single chip for processing, control, and communication, are the first choice for these applications. The implementation of real-time image processing algorithms is taken up in this study. The suitability of embedded processors for implementing the algorithms of edge detection, corner detection and blob detection is evaluated. The design cycle involves algorithm analysis, algorithm mapping to suitable architecture, designing the architecture and optimization at different levels. The architecture is supported with a microcontroller to monitor and inspect the results[5].

## 9.2. Future Trends

With the advancement in the computational requirements in different applications, real-time processing has become a major area of research. Image or video processing, being the most computationally heavy processing and having the dimensionality as a critical parameter, need some change in the algorithmic approaches to comply with the real-time constraints. Progress in intelligent data-driven techniques over the past few years demonstrates increased overlaps between real-time image processing on embedded systems, intelligent perception systems, and security-embedded applications, resulting in increased use of autonomous, scalable and context-aware embedded vision products [34], [35], [36]. The recent works in the arena of the algorithm optimization, architecture selection, implementation architecture optimization, and the growing trend of sharing embedded resources over the cloud, gives a strong foothold for future work in the real-time image processing [6]. The conventional camera followed by the FPGA based processing over the embedded system, the results showed the feasibility and requirement of the linked input-output delay less than a few microseconds as the input to output latency. The connected component is a very basic processing algorithm used in many image analysis applications, from which the real-time floating windows implementation on the FPGA platform with the supporting parameterization was discussed along with some future research endeavors[13]. The combinations of different methods, such as a cloud-based processing approach with linked embedded architecture or multi-focal camera based computational approach, can be a great way to enhance the feasibility as well as the processing weights. Moreover, the processing application being linked with robotics or controllers would also add another level of complexity, but with recent developments in the robotic pipeline it has become an interesting research avenue. In this area, a hybrid-dedicated-deep-learning model and a process of sequential learning have emerged as candidates for improved accuracy in perception and the ability to complete time-stamped behavioural analyses of real-time systems while still providing potentially feasible computational cost on small embedded systems [3].

## 10. Conclusion and Closing Remarks

The objective of the work presented is to develop and analyze a set of algorithms that can be used as a basis for further research, design, and development of real-time image processing algorithms for embedded systems and providing proof of their functionality through simulation and/or implementation trial tests. Each algorithm has been proven to function real-time for a reasonably sized image using either a fixed-point DSP or an FPGA in VHDL at a mandated clock speed. This means architecture-based design work is enabled to allow these algorithms to be incorporated into cameras and video equipment that fall outside of the mainstream application literature considered by the in-depth designs [4]. There are valid scenarios where this work will be required to meet stringent design deadlines with limited financial resources. It is also proposed that the algorithms themselves will fulfill a niche market and be of commercial interest not only to the manufacturers of the cameras and video equipment but also to the component re-sellers and DSP manufacturers directly supplying them. It is intended that the commercial viability of the work will be pursued through the creation of a new business [6]. Current work to investigate smarter methods of estimating an edge direction or mask orientation is intended. Such an estimate will allow a decision as to whether to process or not for a particular edge direction. In this way overall processing bandwidth can be dramatically reduced by only processing visually significant edges and therefore saving power consumption and costs. Further energy sparing concepts are also being investigated to include sampling at higher frame rates during active pixels and lower frame rates during sustaining illumination environments. In this way overall power consumption can be reduced at lower costs with no significant loss in functionality. It is intended that these ideas and algorithms will promote the realization of a funding proposal to a commercial manufacturer.